# An Algorithm to Improve Performance over Multihop Wireless Mesh Network

Ms. Sumedha R. Chokhandre, Ms. Urmila Shrawankar

**Abstract**— Transmission Control Protocol (TCP) is the dominant reliable transport protocol utilized in the Internet. Improving the performance of TCP associated with the presence of multi-hop is one of the research challenges in wireless mesh networks. Wireless mesh networks have large round trip time variations and these variations are dependent on the number of hops. In wireless mesh network, when congestion loss and wireless loss are co-existed the number of packets dropped increases and will have adverse effects on TCP and its congestion control mechanism which leads to low throughput. Here we have designed a new TCP scheme for multi-hop wireless mesh networks, by modifying the sender side congestion control functionality of TCP NewReno, which is tuned towards improving the performance of TCP. The simulation results show that TCP SAC has higher performance than TCP NewReno, Reno, Sack and Vegas in multi-hop wireless mesh networks.

——————————— ◆ ———————————

## 1 INTRODUCTION

WIRELESS mesh networks (WMN) are rapidly emerging as a promising complement to existing broadband access infrastructures, because they extend wireless local area network (WLAN) access beyond traditional hotspot areas to enhance coverage and provide seamless mobility. In WMNs, all communications between mesh network nodes are over radio links. Applications of WMNs include backhaul for broadband access networks, metropolitan area mobile networks, and citywide surveillance systems.

TCP (Transmission Control Protocol) was designed to provide reliable end-to-end delivery of data over unreliable networks. In theory, TCP should be independent of the technology of the underlying infrastructure. In particular, TCP should not care whether the Internet Protocol (IP) is running over wired or wireless connections. In practice, it does matter because most TCP deployments have been carefully designed based on assumptions that are specific to wired networks. Ignoring the properties of wireless transmission can lead to TCP implementations with poor performance. In wireless networks, the principal problem of TCP lies in performing congestion control in case of losses that are not induced by network congestion. Since bit error rates are very low in wired networks, nearly all TCP versions now a day's assume that packets losses are due to congestion. Consequently, when a packet is detected to be lost, either by timeout or by multiple duplicated ACKs, TCP slows down the sending rate by adjusting its congestion window. Unfortunately, wireless networks suffer from several types of losses that are not related to congestion, making TCP not adapted to this environment. Numerous enhancements and optimizations have been proposed over the last few years to improve TCP performance over one-hop wireless (not necessarily Wireless) networks.

The rest of the paper is organized as follows: In Section 2 we describe the Challenges in wireless networks. Section 3 describes the related work for TCP Performance in wireless environment and Section 4 & 5 point out the Proposed Work and Simulation Result. Section 6 & 7 includes conclusion and the references.

## 2 TCP'S CHALLENGES IN WIRELESS NETWORKS

The performance of TCP degrades in wireless networks. This is because TCP has to face new challenges due to several reasons specific to these networks: lossy channels, hidden and exposed stations, path asymmetry, network partitions, route failures, and power constraints.

### 2.1 Lossy Channels
The main causes of errors in wireless channel are the following:
i) **Signal Attenuation:** This is due to a decrease in the intensity of the electromagnetic energy at the receiver (e.g. due to long distance), which leads to low signal-to-noise ratio (SNR).
ii) **Doppler Shift:** This is due to the relative velocities of the transmitter and the receiver. Doppler shift causes frequency shifts in the arriving signal, thereby complicating the successful reception of the signal.
iii) **Multipath Fading:** Electromagnetic waves reflecting

————————————————
• Ms. Sumedha R. Chokhandre is a student of G.H. Raisoni College of Engineering, Nagpur, India.
• Ms. Urmila Shrawankar is a Assistant Professor in Department of Computer Science & Engineering at G.H. Raisoni College of Engineering, Nagpur, India.

off objects or diffracting around objects can result in the signal traveling over multiple paths from the transmitter to the receiver. Multipath propagation can lead to fluctuations in the amplitude, phase, and geographical angle of the signal received at a receiver.

### 2.2 Hidden and Exposed Stations

In wireless networks, stations may rely on physical carrier-sensing mechanism to determine idle channel, such as in the IEEE 802.11 DCF function. This sensing mechanism does not solve completely the *hidden station* and the *exposed station* problems [1]. Before explaining these problems, we need to clarify the "transmission range" term. The transmission range is the range, with respect to the transmitting station, within which a transmitted packet can be successfully received.

A typical hidden terminal situation is depicted in Figure 1. Stations A and C have a frame to transmit to station B. Station A cannot detect C's transmission because it is outside the transmission range of C. Station C (resp. A) is therefore "hidden" to station A (resp. C). Since A and C transmission areas are not disjoint, there will be packet collisions at B. These collisions make the transmission from A and C toward B problematic. To alleviate the hidden station problem, virtual carrier sensing has been introduced [2], [3]. It is based on a two-way handshaking that precedes data transmission.

Specifically, the source station transmits a short control frame, called Request-To-Send (RTS), to the destination station. Upon receiving the RTS frame, the destination station replies by a Clear-To-Send (CTS) frame, indicating that it is ready to receive the data frame. Both RTS and CTS frames contain the total duration of the data transmission. All stations receiving either RTS or CTS will keep silent during the data transmission period (e.g. station C in Figure 1).

However, as pointed out in [4], [5] the hidden station problem may persist in IEEE 802.11 wireless networks even with the use of the RTS/CTS handshake. This is due to the fact that the power needed for interrupting a packet reception is much lower than that of delivering a packet successfully. In other words, node's transmission range is smaller than the sensing node range.

The exposed station problem results from a situation where a transmission has to be delayed because of the transmission between two other stations within the sender's transmission range. In another case, we show a typical 5 scenario where the exposed terminal problem occurs. Let us assume that A and C are within B's transmission range, and A is outside C's transmission range. Let us also assume that B is transmitting to A, and C has a frame to be transmitted to D. According to the carrier sense mechanism, C senses a busy channel because of B's transmission.

Therefore, station C will refrain from transmitting to D, although this transmission would not cause interference at A. The exposed station problem may thus result in a reduction of channel utilization. It is worth noting that hidden terminal and exposed terminal problems are correlated with the transmission range. By increasing the transmission range, the hidden terminal problem occurs less frequently. On the other hand, the exposed terminal problem becomes more important as the transmission range identifies the area affected by a single transmission.

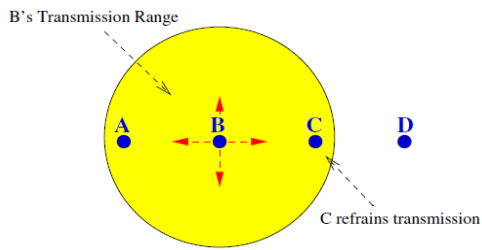

Figure1. Hidden terminal problem: Packets sent to B by A and C will collide at B.

### 2.3 Path Asymmetry

Path asymmetry in wireless networks may appear in several forms like bandwidth asymmetry, loss rate asymmetry, and route asymmetry.

**Bandwidth asymmetry:** Satellite networks suffer from high bandwidth asymmetry, resulting from various engineering the satellite-earth link over the bandwidth of the earth-satellite link is about 1000 [6]. For example, the bandwidth ratio lies between 2 and 54 in Wireless networks that implement the IEEE 802.11 version g protocol [2]. The asymmetry results from the use of different transmission rates. Because of this different transmission rates, even symmetric source destination paths may suffer from bandwidth asymmetry.

**Loss rate asymmetry:** This type of asymmetry takes place when the backward path is significantly lossier than the forward path. In Wireless networks, this asymmetry is due to the fact that packet losses depend on local 6 constraints that can vary from place to place. Note that loss rate asymmetry may produce bandwidth asymmetry. For example, in multi-rate IEEE 802.11 protocol versions, senders may use the Auto-Rate-Fallback (ARF) algorithm for transmission rate selection [3]. With ARF, senders attempt to use higher transmission rates after consecutive transmission successes, and revert to lower rates after failures. So, as the loss rate increases the sender will keep using low transmission rates.

The first one is "TCP header compression" that reduces the size of the TCP ACKs on the backward path [7]. The second one is "ACK filtering" that reduces the number of TCP ACKs transmitted, by taking advantage of the fact that TCP ACKs are cumulative [8]. The third one is "ACK congestion control" that let the receiver also control the congestion on the backward path. This is done by dynamically maintaining a delayed-ACK factor $d$ by the receiver and by sending one ACK for every $d$ data packet received [8]. The difference between ACK filtering and ACK congestion control is that the first one is a link layer technique that can be implemented at intermediate nodes, however the second one is a TCP layer technique that is implemented at the TCP sink. Unfortunately, these techniques alone cause problems such as increasing send-er's burst traffic and also slowing down the sender's congestion window growth. So, it is necessary to adapt the sender congestion control algorithm to avoid these problems. For details about the sender adaptation techniques, we refer to [8]. The adaptive delayed-ACK proposed in [9] aims to reduce the contention on the channel, by reducing the number of TCP ACKs transmitted.

### 2.4 Power Constraints

In general, Wireless networks there are two correlated power problems: the first one is "power saving" that aims at reducing the power consumption; the second one is "power control" that aims at adjusting the transmission power of mobile nodes. Powers saving strategies have been investigated at several levels of a mobile device including the physical layer transmissions, the operation systems, and the applications [11]. Power control can be jointly used with routing or transport agents to improve the performance of Wireless networks [12], [13]; power constraints communications reveal also the problem of cooperation between nodes, as nodes may not participate in routing and forwarding procedures in order to save battery power.

## 3 RELATED WORK

Several efforts for improving the performance of TCP in multi-hop wireless networks have recently been reported. The problem of TCP performance and degradation over multi-hop wireless network due to the conflict between data packets and acknowledgments are identified in [14]. Also because of the impact of the MAC protocol on performance of TCP in multi-hop networks. The TCP throughput decreases exponentially as the number of hop increases due to hidden terminal problem, which increases the packet collision. Similar problems were evaluated in [15] where the authors show that using smaller values for both packet size and maximum window size in TCP setup can mitigate such problems to some extent. Several implementations of TCP are analyzed in [16] and they discovered that the throughput of TCP depends on the number of hops in the path as well as the performance of underlying routing protocol.

An end to-end combination scheme is evaluated in [17] to improve TCP throughput over multi-hop wireless network. In literature, different approaches focused on improving the throughput of TCP over multi-hop wireless networks have been described. Although much research has been done on improving TCP performance, only few approaches have been proposed for improving TCP performance in wireless mesh networks. A multi-channel assignment algorithm is proposed in [18] for constructing a wireless mesh network which eliminates the hidden terminal problem and thereby improving the performance of TCP in wireless multi-hop networks. A pacing scheme is proposed in [19] at the IP layer to improve the fairness in hybrid wireless/ wired networks. Gambiroza, Sadeghi and Edward [20][21][22] studied TCP performance in wireless mesh networks and they proposed a distributed link layer algorithm for achieving fairness among TCP flows. These approaches do not have a mechanism to avoid retransmission timeout caused by retransmission loss. Our proposed algorithm constitutes a modification at the transport layer in sender side rather than a modification at the link layer for improving the performance of TCP by avoiding the frequent retransmission timeout in multi-hop wireless mesh networks [23].

## 4 PROPOSED WORK

To conquer the throughput degradation of TCP in multihop wireless mesh networks, we modified the fast retransmit and recovery algorithms of TCP NewReno. The key idea of our algorithm is to calculate the outstanding packets and set the value of slow start threshold (ssthresh) based on half of the difference between maximum data packets sent and the last acknowledgment received at the TCP sender. We used this difference for minimizing the frequent retransmission timeouts caused by retransmission loss. The value of ssthresh is used to determine whether the TCP should do Slow Start or Congestion Avoidance. TCP SAC adopts Slow Start and Congestion Avoidance of TCP NewReno and modifies Fast Retransmit and Recovery algorithms.

Figure 3 shows the phase transition diagram of TCP SAC. At the beginning of a connection, congestion window (cwnd) is initialized to maximum segment size (t_mss) and ssthresh is set to an arbitrarily large value, 65535 bytes. The variable ssthresh is maintained to determine the state of a TCP sender.

**Slow Start (SS):**

In Slow Start state (SS), when the new connection is established the cwnd is initialized to one packet. Every time a packet with sequence number (n) arrives at the receiver, the receiver sends an acknowledgment to the sender. After receiving the acknowledgment, the size of the cwnd is set from one to two, and two packets can send to the receiver.

In the case where each of those two packets receives correctly, the size of cwnd is increased to four. This can be estimated as an exponential growth. This process continues until the sender receives three duplicate acknowledgments or retransmission timeout. When congestion window size is greater than ssthresh value, the sender moves to Congestion Avoidance state from Slow Start State.

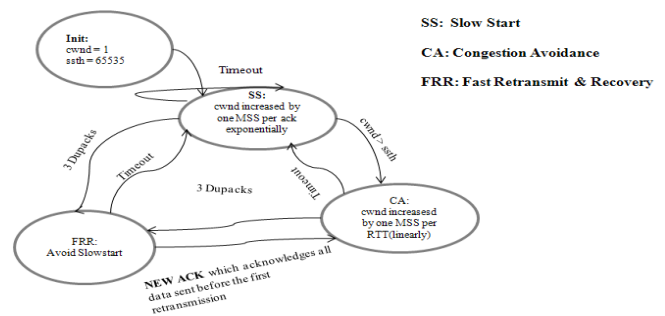

Figure 3: State diagram of TCP SAC

**Congestion Avoidance (CA):**

In Congestion Avoidance state, cwnd is increased linearly so that the sender can probe the network and increase the sending rate gradually without causing congestion too frequently.

**Fast Retransmit and Recovery (FRR):**

In this state, whenever a packet is received out of order, the receiver sends duplicate acknowledgments (t_dupacks) with the same sequence number to the sender. When the number of duplicate acknowledgments reaches the threshold value (tcpxmthresh) which is equal to three, sender moves to FRR state. As a result, TCP sender invokes Fast Retransmit algorithm and stores the highest sequence number sent to the variable High_seq. The sender calculates the number of outstanding packets (t_nps), and reduces the value of ssthresh. After that, the sender retransmits the lost packet and reduces the cwnd based on the value of ssthresh plus three times to the segment size. This inflates the congestion window by the number of segments that have left the network and the receiver has buffered. After inflating cwnd, the sender stores the retransmission loss point in the variable Rlp. Then TCP enters into Fast Recovery algorithm it count the additional duplicate acknowledgments (add_dupacks). These duplicate acknowledgments indicate the reception of some segments by the receiver. For each additional duplicate acknowledgment, sender increases its cwnd by one segment size and transmits segments allowed by the new value of
cwnd.

If the duplicate acknowledgments are greater than or equal to the stored value of variable Rlp minus one (here minus one represents the lost packet via duplicate acknowledgments), then the sender detects the retransmitted packet is lost. Without waiting for timeout expiration, the sender retransmits the lost packet immediately and reduces the cwnd to half of the present value of cwnd (p_cwnd). But the value of ssthresh remains the same. When an acknowledgment arrives that acknowledges new data including the stored value of the variable High_seq, then the TCP sender exits Fast Recovery algorithm and goes to CA state by setting the value of cwnd equals to ssthresh. In this way, we saved the TCP from frequent retransmission timeout caused by retransmission loss and can increase the throughput of TCP in multi-hop wireless mesh networks.

## 5 SIMULATION RESULTS

This section represents the evaluation results of TCP SAC compare with other key existing TCP flavors such as NewReno, Reno, Sack and Vegas. Throughputs, Delay & Loss are the main performance metrics we used in the evaluation of TCP SAC using the network simulator 2.

### 5.1 Throughput Performance

The throughput is calculated based on the total number of data packets sent divided by the time difference between the last packet sent and the first packet sent.

We compared the throughput of TCP SAC with TCP NewReno, Reno, Sack and Vegas on various network loads using the chain topology as shown in figure 3. For causing retransmission loss we adjusted the time of exponential error model. In addition to throughput evaluation, we calculated the Packet Loss Rate (PLR) at the transport layer. It is defined as the number of data packets retransmitted divided by the number of data packets received. We conducted the simulation in the following ways: In each simulation we set node 1 as the source node and we changed the destination according to the number of hops as shown in Figure 4. Figure 5, 6 and 7 shows the throughput of TCP SAC compared to TCP NewReno, Reno, Sack and Vegas in different Packet Loss Rates and network loads.

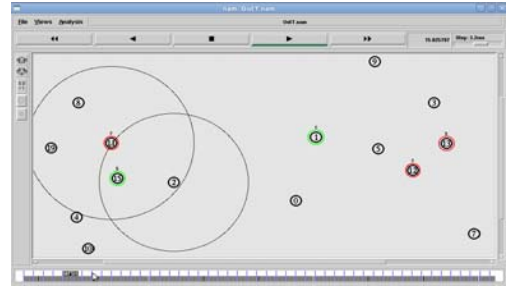

Figure 4. Transmission of Data form various nodes

When the packet loss rate increases throughput decreases with the number of hops (here we used four hops for our evaluation). This means that, RTT is increasing according to number of hops and leads to increase the number of timeouts. As a result, TCP NewReno, Reno and Sack cut the congestion window size to 1MSS and goes to Slow Start state. In the case of Vegas it can avoid retransmission timeout but it cannot avoid the timeout due to retransmission loss like other TCP schemes.

This reduces the sending rate quickly and causes throughput degradation. But in the case of TCP SAC, it does not wait for retransmission timeout. Instead of that, it retransmits the packet caused by retransmission loss in FRR state itself and not reducing the value of ssthresh on each loss. In Figure 5, 6 & 7 we observe that the throughput of TCP SAC has significant improvement in the performance compared to other TCP schemes.

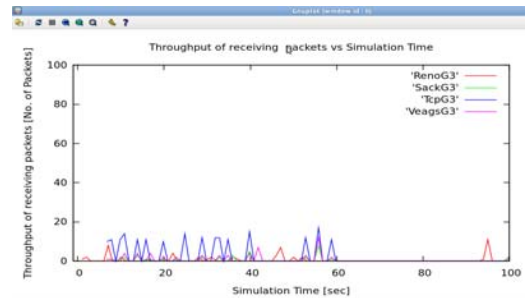

Figure 5:- Throughput of receiving packets

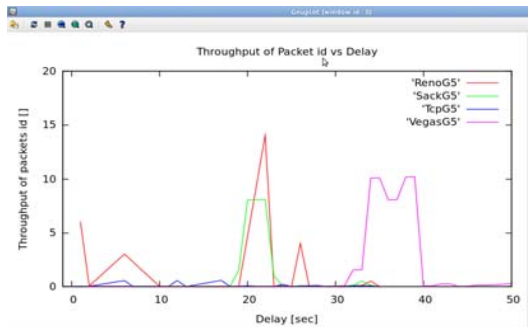

Figure 6:- Throughput of Delay

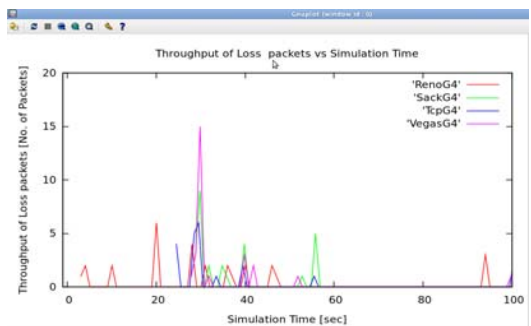

Figure 7:- Throughput of Packet Loss

## 6 CONCLUSION

We evaluated the performance of TCP SAC over multi-hop wireless mesh networks through extensive simulations using Network Simulation-2. The simulation results have confirmed that TCP SAC has a significant improvement over TCP NewReno. Two salient features of TCP SAC contribute to this improvement. First, reduced the retransmission timeout caused by retransmission loss and second, adjusted the size of ssthresh and cwnd in Fast Retransmit and Recovery algorithms. These two features help TCP SAC to increase the performance of TCP over multi-hop wireless mesh networks.

**Ms. Sumedha R. Chokhandre** is a student IV Semester M.E.   (Wireless Communication & Computing) at G.H.Raisoni College of Engineering, Nagpur, India.

**Ms. Urmila Shrawankar** is an Assistant Professor, Department of Computer Science & Engineering at G.H.Raisoni College of        Engineering, Nagpur, India.